\documentclass[epj]{svjour}
\usepackage{graphicx}
\usepackage{amsmath}
\usepackage{cancel}
\begin{document}
\title{Fractional hyperviscosity induced growth of bottlenecks in energy
spectrum of Burgers equation solutions} \author{Debarghya Banerjee}
\institute{Max Planck Institute for Dynamics and Self-Organization, Am Fa\ss berg
17, G\"ottingen 37077 Germany}
\date{\today}
\abstract{ Energy spectrum of turbulent fluids exhibit a bump at an
intermediate wavenumber, between the inertial and the dissipation range. This bump is called
bottleneck. Such bottlenecks are also seen in the energy spectrum of the
solutions of hyperviscous Burgers equation. Previous work, have shown that this
bump corresponds to oscillations in real space velocity field. In this paper we
present numerical and analytical results of how the bottleneck and its'
real space signature, the oscillations, grow as we tune the order of
hyperviscosity. We look at a parameter regime $\alpha \in [1,2]$ where
$\alpha = 1$ corresponds to normal viscosity and $\alpha = 2$ corresponds
to hyperviscosity of order 2. We show that even for the slightest fractional increment in the
order of hyperviscosity ($\alpha$) bottlenecks show up in the energy spectrum.
\PACS{
      {47.10.-g}{Fluids}   \and
      {05.45.-a}{Non-linear dynamics}
     } % end of PACS codes
} %end of abstract
\titlerunning{Fractional hyperviscosity}

\maketitle
\section{Introduction}
\label{intro}

High resolution direct numerical simulations and experimental data have shown
the presence of bottleneck in the kinetic energy spectrum
~\cite{Kaneda2003,Dobler2003,Beresnyak2011,Donzis2010}. 
The energy spectrum is defined as $E(k) \equiv 1/\Delta \Sigma_{{\bf k} \in [k,k+\Delta]} {\bf v}({\bf k})^2$,
where, ${\bf v}$ is the velocity field of the turbulent fluid.
Typical fluid turbulence energy spectrum
is characterised by two length scales -- the large length scale
at which energy is injected in the flow, the corresponding wave
number being $k_{\rm inj}$, and the small length
scale at which the energy dissipation due to viscosity becomes important and
corresponds to a wavenumber $k_{\rm diss}$. Now, in between these wavenumbers
we have the inertial range and is characterised by a scaling behaviour in the
energy spectrum i.e. for $k_{\rm inj} \ll k \ll k_{\rm diss}$ we have $E(k)
\sim k^{-5/3}$ where the exponent $5/3$ corresponds to the Kolmogorov
exponent~\cite{Kolmogorov1941,Frisch1995}.  In the solutions of Burgers equation 
one can also define an inertial range but with a scaling behaviour given
by $E(k) \sim k^{-2}$. Bottleneck is characterised by the
presence of a bump at an \emph{intermediate} wavenumber, between the inertial
and the dissipation wavenumbers. 
The bump in this intermediate wavenumber
occurs due to an inefficient transfer of energy across
wavenumbers~\cite{Falkovich1994} and manifests as oscillation in the real
space correlation functions~\cite{Frisch2013}. Another aspect of bottlenecks
that has been discussed in literature before is the fact that the effect of
bottlenecks gets more pronounced if one uses hyperviscosity instead of
viscosity. To get a clearer theoretical understanding of the bottlenecks one
needs to consider the one dimensional Burgers
equation~\cite{Burgers1948,Bec2007,Bec2002,Frisch2002} in the presence of
hyperviscosity. It has been shown in Ref.~\cite{Frisch2013} that oscillations
emerge in the solutions of Burgers equation in the presence of hyperviscosity
which can be causally related to the bottlenecks in the energy spectrum.
However, unlike fluid turbulence it can also be shown that these oscillations
vanish in the limit of a normal viscous Burgers
equation~\cite{Frisch2013,Frisch2008,Banerjee2014}.

The use of hyperviscosity, to introduce strong dissipation at large wavenumbers
leading to an extended inertial range behaviour, have been a common practice in
direct numerical simulations of turbulence. In the usage of hyperviscosity one
introduces a linear dissipative term in the Navier-Stokes equation either
replacing the normal viscous term or in conjunction to the normal
viscosity~\cite{Erland2004}. The hyperviscous dissipation has the form of $\nu_{\alpha}
(-1)^{\alpha+1} (\nabla^2)^{\alpha} {\bf v}$, where $\alpha$ is the order of
hyperviscosity, $\nu_{\alpha}$ is the coefficient of hyperviscosity, and ${\bf
v}$ is the velocity field. In fact $\alpha=1$ corresponds to normal viscosity
and $\alpha > 1$ corresponds to hyperviscosity. The presence of hyperviscosity
is known to lead to a wide range of interesting phenomenon both in turbulence
and in the simpler and more tractable one dimensional Burgers equations. 

Very high values of $\alpha$ (hyperviscosity) leads to
a thermalised solution of the Burgers equation~\cite{Frisch2008,Banerjee2014}.
Thermalisation or equipartition
of energy among the wavenumber modes was known to happen for the Galerkin
truncated Euler and inviscid Burgers equation~\cite{Cichowlas2005}. The
combined operation of making viscosity 0 and Galerkin truncation makes the equations conservative with a finite number of
degrees of freedom. Finite dimensional conservative system are known to thermalise.  As we increase the order
of hyperviscosity in the Burgers equation we see the generation of bottlenecks
in the energy spectrum whose real space signature corresponds to oscillatory
structures at the shock front. On increasing the order of hyperviscosity the
oscillations become more and more delocalised. At some point the oscillations
reach the stagnation point and forms \emph{tygers}~\cite{Ray2011,Ray2015}; and finally the localised
structure at the stagnation point spreads to all over the system leading to
thermalisation~\cite{Banerjee2014}. 

A very high order of hyperviscosity essentially mimics a Galerkin truncation
and using theoretical understanding of the Burgers equation it is possible to
locate an estimate of the crossover value of the order of hyperviscosity where
the system goes from a dissipative hyperviscous system to a conservative
thermalised system where equilibrium statistical mechanics work exactly.
The interplay of linear and non-linear terms in the transition from dissipative
to conservative systems have been an important topic of scientific research
in the recent past~\cite{Thalabard2016,Feng2017,Shukla2013}.

As pointed out before for normal viscosity the theoretical prediction is that
there is no bottleneck in the energy spectrum.
Also it is known that for even the lowest integral order of hyperviscosity there is a
finite bottleneck. In this paper we consider fractional order of hyperviscosity
between normal viscosity and the lowest order of hyperviscosity, i.e. the regime $\alpha \in [1,2]$, to see the
development of bottlenecks for small orders of hyperviscosity. We find that
unlike the crossover between dissipative hyperviscous to conservative burgers
the normal viscosity behaves like a critical point for the existence of
bottlenecks and use techniques of fractional calculus to justify our numerical results~\cite{Mainardi1996}.

\section{Numerical setup}
\label{sec:1}

The one dimensional Burgers equation with hyperviscosity can be written as:
\begin{align} 
\frac{\partial v}{\partial t} + \frac{\partial}{\partial x} \frac{v^2}{2} = -\nu_{\alpha} (-1)^{-\alpha} \left(\frac{1}{k_d} \frac{\partial}{\partial x} \right)^{2 \alpha} v + f.  
\label{eq:burg}
\end{align} 
Where, $\nu_{\alpha}$ is the hyperviscosity, $k_d$ is the effective
wavenumber where the hyperviscosity becomes important, and $\alpha$ is the
order of hyperviscosity, $\alpha$ is defined in such a way that $\alpha=1$
corresponds to normal viscosity and $\alpha=2$ corresponds to hyperviscosity of
the lowest integral order. For our purpose $\nu_{\alpha} = 1000$, $k_d =1000$.
The forcing $f$ is chosen as $f = f_{\rm amp} \sin(x)$ for all of our
simulations $f_{\rm amp} = 1$.  Note, that the above choice of $\nu_{\alpha}$
and $k_d$ gives us a very small value of effective hyperviscosity that can be defined as
$\nu_{\alpha}^{\rm eff} = \nu_{\alpha}/{k_d}^{2 \alpha}$.
We solve the Burgers equation numerically by
the standard pseudo-spectral method. Since we solve the equations in Fourier
space the fractional derivatives can be calculated as
$ - \nu_{\alpha} (k/k_d)^{2 \alpha} \hat{v}$.  The system size is $2\pi$ and there
are 16384 grid points.  The choice of forcing $f = \sin(x)$ results in a shock
at $x=\pi$. The oscillatory behaviour at the shock front is extracted by
deducting the solution of a normal viscous Burgers equation from the solution
of the hyperviscous Burgers equation ($v^{\alpha} (x) - v^{1}(x)$).

\section{Result}
\label{sec:2}

\subsection{Numerical results}

In Fig. 1(a) we see the compensated energy spectrum of the solutions of
Eq.~(\ref{eq:burg}) and the corresponding flow structure in real space is given
in Fig.  1(b). The $k^{-2}$ spectrum arises from the Fourier transform of the 
real space shock structure. The functional from of the shock
structure that we see in the Fig. 1(b) can be calculated by solving the
Eq.~(\ref{eq:burg}) using methods from boundary layer techniques at the steady
state. In this method we consider the \emph{outer} solution and the
\emph{inner} solution of the equation. The outer solution gives the large scale
behaviour determined by the large scale forcing and setting viscous term to
zero. While the inner solution corresponds to the effect of viscous term at the
shock front. The oscillations that we see straddling the shock front,
particularly visible on zooming at the shock front in Fig.~\ref{fig:2} (a), is
generated by the effect of the hyperviscosity. Clearly, the viscous ($\alpha =
1$) limit shows no oscillations and hence no bottlenecks while for
hyperviscosity ($\alpha > 1$) we see the oscillations emanating out of the
solutions of Eq.~(\ref{eq:burg}). The analytical solutions that we can
calculate for the integral-hyperviscous cases do not work for the fractional
hyperviscous case but one can still study these cases via simulations using
pseudo-spectral method.

In Fig.~\ref{fig:2} (a) we show how the amplitude of oscillations grow with
increasing $\alpha$. The amplitude of the first visible peak versus $\alpha$ is
fitted using a functional form $a_1 (\alpha - 1)^{a_2}$, where $a_1$ and $a_2$
are the fitting parameters. The fitted function is plotted with a black dashed
line in Fig.~\ref{fig:2} (b) we find that the fitted parameters are $a_1 \sim
0.5$ and $a_2 \sim 0.6$. The oscillation amplitude has a critical behaviour at
$\alpha = 1$ For values of $\alpha$ even fractionally larger than 1 we see the
oscillations developing at the shock front while for $\alpha \le 1$ no such
oscillations develop at the shock front. We later show that this critical
behaviour is due to the presence of singularity leading to non analytic
behaviour. However, it must be noted that reduction of $\alpha$ cannot be
continued too much below 1 as there is another critical point at $\alpha = 0.5$
~\cite{Bardos1979,Miao2009}.  

\subsection{Asymptotic method}

The real space structure that we get from the numerical simulations can also
be obtained by solving Eq.~(\ref{eq:burg}) using matched asymptotes. We need
to essentially match the solutions of the equation obtained at the outer limit
and the inner limit. The first step to solving the equations is to set $\partial_t v = 0$
at the steady state. Now, we are left with an ordinary differential equation.
Next, we consider the outer
limit of the solution and then the inner limit of the solution and a simple
matching would give the complete solution of the equations. To obtain the outer
solution, which corresponds to the large scale behaviour and is unaffected by hyperviscosity,
 we set $\nu_{\alpha}^{\rm eff} = 0$ and hence obtain the equation:
\begin{align}
\frac{\partial}{\partial x} v^2(x)= 2 \sin(x).
\label{eq:outer}
\end{align}
On solving the above equation we
get $v(x) = 2 {\rm sgn}(x - \pi) \sin(x/2)$. Where $x=\pi$ gives the location
of the shock for the forcing used.

The inner solution which essentially gives the inner structure of the shock
due to the effect of the viscous term can be obtained by first rescaling
$X = (x - \pi)/\nu_{\alpha}^{\beta}$, where $\beta = 1/ (2 \alpha -1)$ and then
expanding the velocity in powers of $\nu_{\alpha}^{\rm eff}$. We then assume the limit $\nu_{\alpha}^{\rm eff} \rightarrow 0$
and retain the leading order term.
This leads to the inner equation:
\begin{align}
\frac{d}{d X} v^2(X)= -2 (-1)^{-\alpha} \frac{d^{2\alpha}}{dX^{2\alpha}} v(X)
\label{eq:inner}
\end{align}
The above equations need to solved using boundary conditions that
are a consequence of the outer solution discussed in the previous paragraph
which is $v(\pm \infty) = \mp 2$.
Now, considering the
$\alpha =1$ case we are left with:
\begin{align}
\frac{d}{d X} v(X) = \frac{1}{2}(v^2(X) - 4),
\end{align}
on solving we get $v(X) = -2 \tanh(X)$. For the hyperviscous case of $\alpha =2$
the solution is a bit more tricky to get and we need to consider the linear disturbances
from a zeroth order shock solution. To do so we write $v(X) = -2 + w(X)$, for $X > 0$
and retain terms linear in $w(X)$. This gives us
a linear ordinary differential equation
in $w(X)$ where $w(X)$ is the linear disturbance, hence we have:
\begin{align}
\frac{d^3}{d X^3} w(X) = 2 w(X)
\end{align}
To solve the above equation we consider solutions of type $e^{\mu X}$ and plug it in the
above equation this gives us the algebraic equation $\mu^3 - 2 = 0$. Now, $\mu$ has a real solution and
a pair of complex conjugate solutions. The complex solutions for $\mu$
gives us oscillatory behaviour which characterises the
oscillations straddling the shock and hence leading to bottlenecks in energy spectrum.
For $\alpha = 1 + \varepsilon$ we need to use techniques of fractional calculus to solve
the equation for fractional $\varepsilon$.
Interestingly though, there is an intermediate $\alpha = 3/2$ where the differential equation
is non-fractional:
\begin{align}
\frac{d^2}{d X^2} w(X) = -2 i w(X) 
\end{align}
The solution of this equation gives oscillatory solutions like that of
the higher integral order $\alpha$ cases.
At this stage let us quickly review the solutions for $\varepsilon = 0$ in the normal $x$- coordinate.
From matching the inner and outer solution we get $v(x) = -2 \tanh(x/2 \nu_{\alpha}^{\rm eff}) \sin(x/2)$.
The $\tanh$--function leads to a spectrum with a scaling $|\hat{v}(k)| \sim 1/k$ in the limit of
$\nu_{\alpha}^{\rm eff} \rightarrow 0$.

\subsection{Fractional derivatives}

In the previous two subsections we have been able to understand using numerical
simulations and asymptotic methods the presence of oscillations around the shock
for hyperviscous burgers equation with integer order of hyperviscosity and also
the case of non-interger hyperviscosity with order $\alpha=3/2$, where the differential equation reduces to
a non-fractional differential equation.
To understand how fractional derivatives lead to the growth of oscillatory
behaviour in the shock front let us start by looking at the linear equation
for the disturbance from a shock solution. Thus we start with:
\begin{align}
- (-1)^{-\alpha} \frac{d^{2 \alpha -1}}{d X^{2 \alpha -1}} w(X) = 2 w(X).
\end{align}
Now, we rewrite the above equation as and discuss the regime where $\alpha \in [1,2]$:
\begin{align}
\frac{d^{\gamma}}{d X^\gamma} w(X) = -2 e^{i \pi \alpha} w(X),
\end{align}
where, $\gamma = 2 \alpha - 1$.
Now we use the properties of fractional derivatives as discussed in the Appendix
to rewrite the differential equation as:
\begin{align}
\frac{d^{\gamma}}{d X^{\gamma}} \left( w(X) - \Sigma_{k=0}^{m-1} \frac{c_k t^k}{k!} \right) = -2 e^{i \pi \alpha} w(X).
\end{align}
here, $m-1$ is the integer part of $\gamma$. Using the above equation and
the Laplace transform method discussed in Appendix we obtain the solutions
in terms of complex integrals as shown below:
\begin{align}
w(X) = \frac{1}{2 \pi i} \int_{Br} e^{s x} \frac{s^{\gamma}}{s^{\gamma} + e^{i \pi \alpha}} ds
\end{align}
Where we are integrating along the Bromowich path which is $Re(s) = \sigma$.
Also the solution can be divided into two parts: $w(x) = w_0(x) + w_1(x)$.
The first part $w_0(x)$ consist of the integration along the outer contour; i.e.:
\begin{align}
w_0(x) = \frac{1}{2 \pi i} \int_{Ha(\epsilon)} e^{s x} \frac{s^{\gamma}}{s^{\gamma}+ e^{i \pi \alpha}} ds
\end{align}
and the second part $w_1(x)$ consist of the residues at the poles present in the
main Reimann sheet which implies:
\begin{align}
w_1(x) = \Sigma_h e^{s_h' x} Res\left[ \frac{s^{\gamma}}{s^{\gamma} + e^{i \pi \alpha}} \right]
\end{align}
Now imposing the condition that the pole has to be present in the first Reimann
sheet gives us the following relation:
\begin{align}
\left| \frac{\pi \alpha }{2 \alpha -1} \right| < \pi
\end{align}
Clearly from the above inequality we can see that 
for $\alpha > 1$ there is a pole in the physical Reimann sheet
and for $\alpha \le 1$ this pole is not present in the 
physical Reimann sheet.  This absence of the pole singularity
leads to an absence of oscillations in the physical velocity field
for $\alpha \le 1$.

\begin{figure*}[htbp!]
\includegraphics[width=0.95\columnwidth]{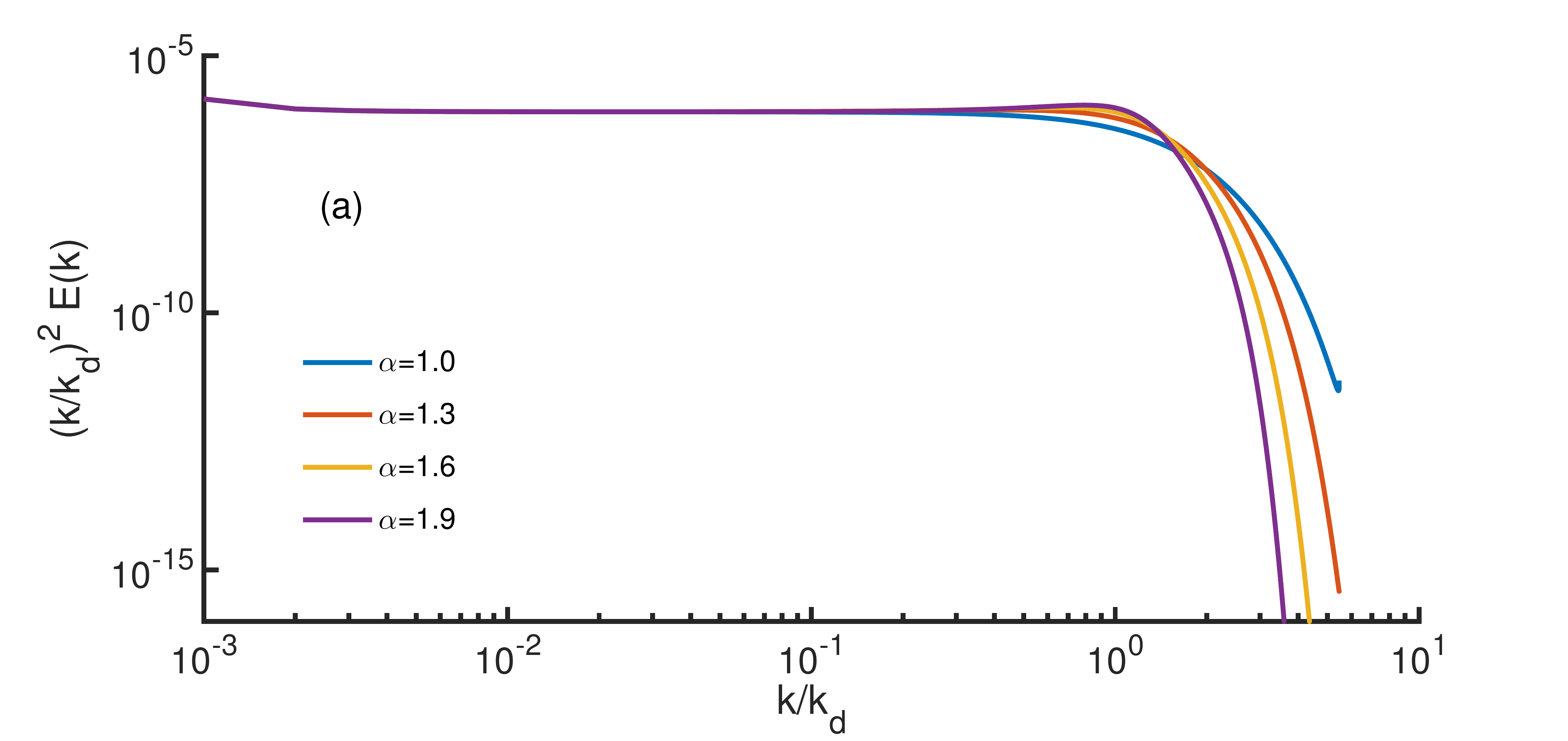}
\includegraphics[width=0.95\columnwidth]{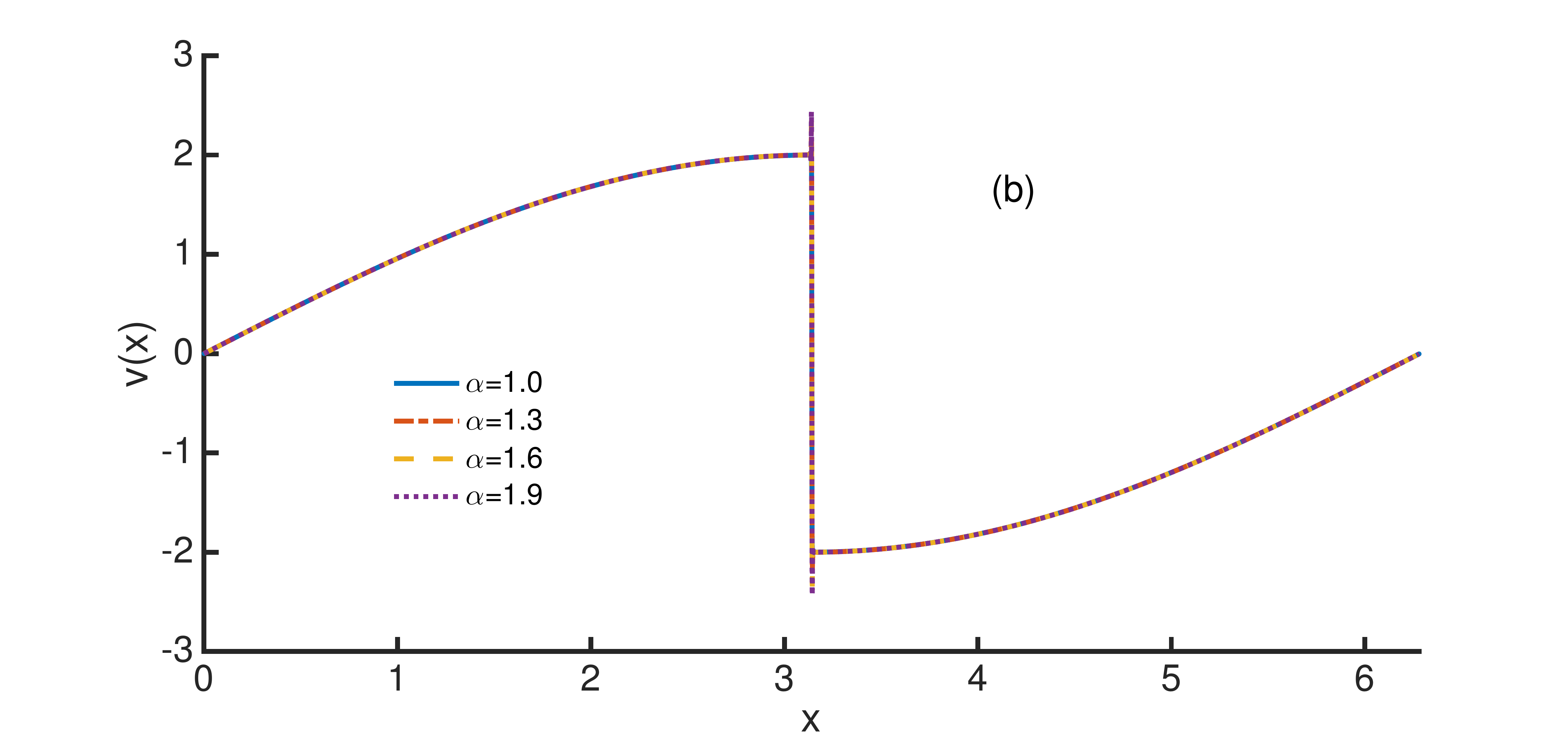}
\caption{Results of direct numerical simulations of the Burgers equation with hyperviscosity such that
$\alpha \in [1,2]$. In (a) we plot the compensated energy spectra i.e. $(k/k_d)^2 E(k)$ versus $k/k_d$ 
and in (b) we plot the velocity field structure in real space.}
\label{fig:1}
\end{figure*}

\begin{figure*}[htbp!]
\includegraphics[width=0.95\columnwidth]{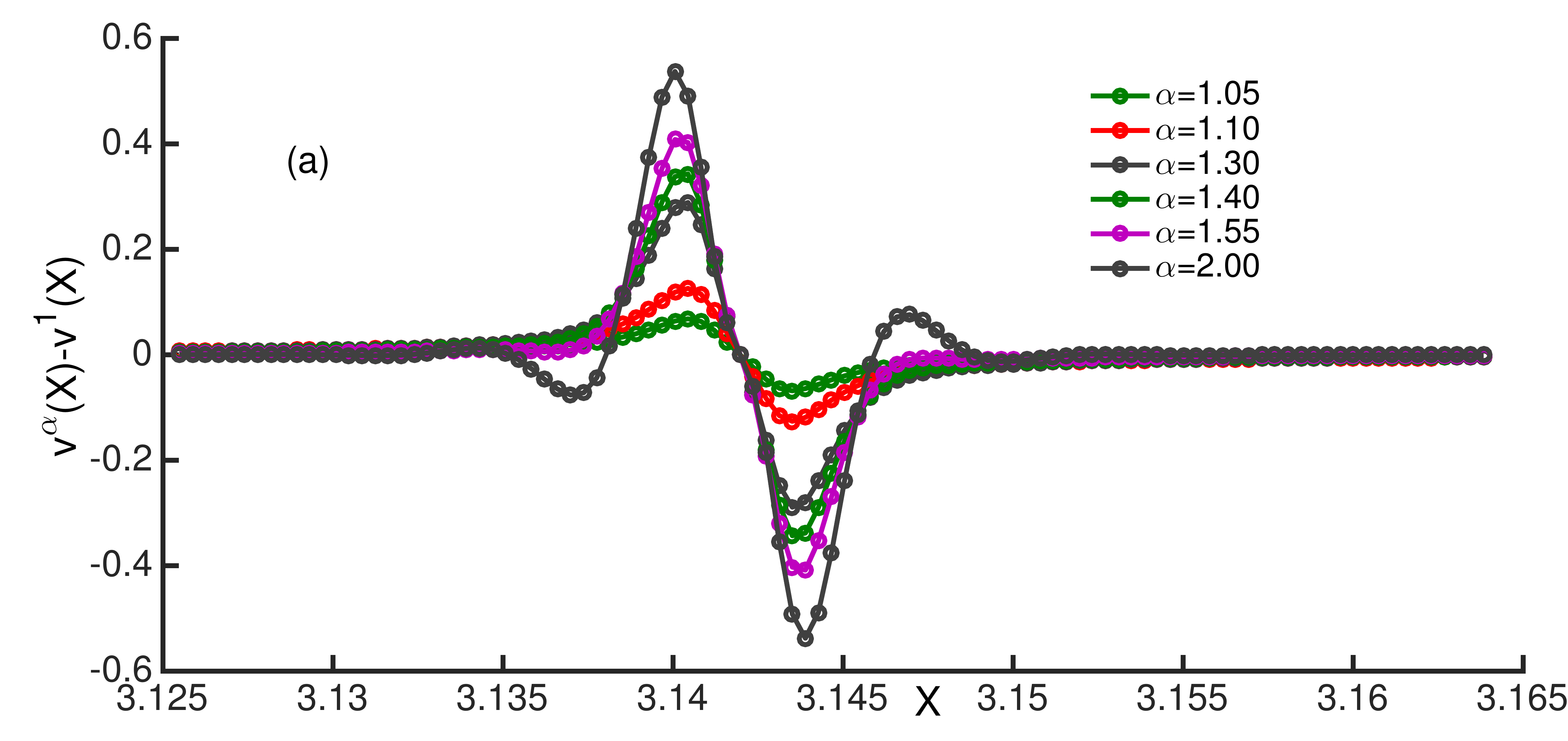}
\includegraphics[width=0.95\columnwidth]{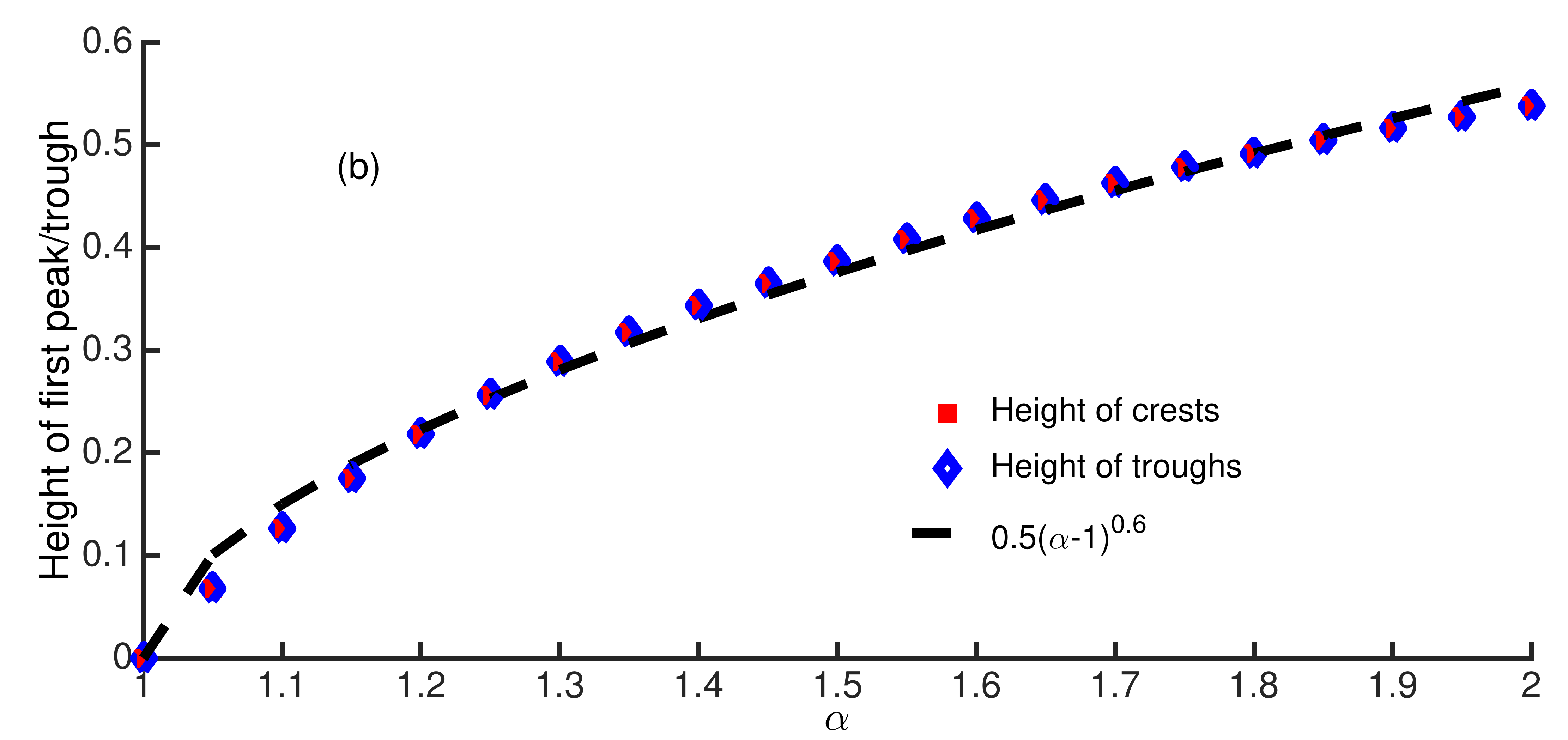}
\caption{ Detailed exploration of the shock structure gives us the above details. 
In (a) we plot the magnitude of discrepancy between the solutions of the hyperviscous
Burgers and the viscous Burgers equations ($|v^{\alpha}(x) - v^1(x)|$ vs. $x$) 
keeping all other parameters like time of measurement, 
initial amplitude, and value of the coefficient of hyper viscosity same. In (b) we show how the 
absolute value of the peak or trough increases as we go from the limit of $\alpha=1$ to $\alpha=2$. 
The dashed line in (b) is obtained by fitting the data.}
\label{fig:2}
\end{figure*}

\section{Summary and conclusions}
\label{sec:3}

Shock front oscillations and as a consequence bottlenecks
are generated in the Burgers equation by hyperviscosity even if it is present
to a very low order.  For integral order of hyperviscosity asymptotic methods
have been employed to show the presence of these oscillations but for fractional
order of hyperviscosity the verification is possible only by direct numerical
simulations. In this paper we consider fractional order of hyperviscosity
between normal viscosity and the lowest order of hyperviscosity to see the
development of bottlenecks for small orders of hyperviscosity. We find that
unlike the crossover between dissipative hyperviscous to conservative burgers
the normal viscosity behaves like a critical point for the existence of
bottlenecks.

Simplified hydrodynamic equations like the Burgers equation form a key basis
for understanding a more complicated problem of fluid turbulence. The work
presented in this paper would open questions more pertinent to the problem of
fluid turbulence as a physical problem and the Navier-Stokes equation as a
mathematical problem. One of the most important questions that arise now is
related to fact that bottlenecks form in normal viscous Navier-Stokes
turbulence but is it possible to have some fractional $\alpha < 1$ such that
there is a transition to a no-bottleneck spectra ? It remains to be
investigated using high resolution direct numerical simulations of
Navier-Stokes equation. Also the next question that would arise is whether such
a transition is critical like the case of Burgers equation.

Hyperviscosity has been used quite commonly in research papers studying
turbulence in fluids and plasmas~\cite{Pandit2017,BanerjeeMHD2014,BanerjeeMHD2019} using both
direct numerical simulations and shell models.  The use of hyperviscosity
becomes absolutely essential for certain class of equations like the Hall
MHD~\cite{BanerjeeHMHD2013} equations where the non linearity has different
dominant behaviour for smaller length scales. The presence of fractional
hyperviscosity in Hall MHD equations can be an important problem to
investigate. Fractional diffusion has been used as an important way to model
diffusion in polymer solutions and the signature of complex singularities on
real space structures may even be probed experimentally.

\section{Acknowledgements}
DB would like to thank U. Frisch, R. Pandit, S. S. Ray, and W. Pauls 
for discussions and COST Action MP 1305 for support.

\section{Appendix}

In this appendix we show how we solve the fraction differential
equations discussed in the results section.

Let us consider the fractional differential equation as discussed in the paper:
\begin{align} 
\frac{d^{\gamma}}{d X^{\gamma}} w(X) = -2 e^{i \pi \alpha} w(X).
\end{align} 
Let us also define operators $D^{\gamma}$ and $J^{\gamma}$ such
that $D^{\gamma}$ is a fractional differential operator and $J^{\gamma}$ is a
fractional integral operator. Amongst the various properties of these operators
what is very useful for us here is the property that $D^{\gamma} J^{\gamma} =
{\bf I}$ but $J^{\gamma} D^{\gamma} \neq {\bf I}$. The detailed explanation can
be found in the literature for example in Ref.~\cite{Gorenflo}.

The above differential equation can therefore be written as:
\begin{align}
D^{\gamma} \left( w(X) - \Sigma_{k=0}^{m-1} \frac{c_k X^k}{k!} \right) = -2 e^{i \pi \alpha} w(X)
\end{align}
On applying the operator $J^{\gamma}$ and taking the Laplace transform we get:
\begin{align}
\tilde{w}(s) = \Sigma_{k=0}^{m-1} \frac{c_k}{s^{k+1}} - 2 e^{i \pi \alpha} \frac{1}{s^{\gamma}}\tilde{w}(s),
\end{align}
which in turn gives us:
\begin{align}
\tilde{w}(s) = \Sigma_{k=0}^{m-1} c_k \frac{s^{\gamma - k - 1}}{s^{\gamma} + e^{i \pi \alpha}}.
\end{align}
To obtain $w(X)$ we need to take a Laplace transform of the above expression $\tilde{w}(s)$.
In this Laplace transform integral, for a pole type singularity to occur we need $s = e^{-i (2 h + 1) \pi \alpha/\gamma}$, with integer $h$.
Now the phase in the exponential lies in the principle Reimann sheet when $|\pi \alpha/ 2\alpha -1|< \pi$. This
happens only for $\alpha > 1$.

% BibTeX users please use
% \bibliographystyle{}
% \bibliography{}
%
% Non-BibTeX users please use

\end{document}